\documentclass{PoS}
\usepackage{amssymb,amsmath,wrapfig}

\title{Spontaneous supersymmetry breaking in the $2d$ ${\cal N}=1$
  Wess-Zumino model}

\ShortTitle{Supersymmetry breaking in the ${\cal N}=1$
  WZ model}

\author{David Baumgartner, Kyle Steinhauer and \speaker{Urs Wenger} \\
        Albert Einstein Center for Fundamental Physics\\
         Institute for Theoretical Physics\\
        University of Bern\\
        Sidlerstrasse 5\\
        CH--3012 Bern\\
        Switzerland\\
        E-mail: \email{baumgart@itp.unibe.ch}, \email{steinhauer@itp.unibe.ch},
        \email{wenger@itp.unibe.ch}}

     \abstract{ We study the phase diagram of the two-dimensional ${\cal N}=1$
        Wess-Zumino model using Wilson fermions and the fermion loop
        formulation. We give a complete non-perturbative
        determination of the ground state structure in the continuum
        and infinite volume limit.  We also present a determination
        of the particle spectrum in the supersymmetric phase, in the
        supersymmetry broken phase and across the supersymmetry breaking phase
        transition. In the supersymmetry broken phase we observe the emergence
        of the Goldstino particle.}

\FullConference{ The XXX International Symposium on Lattice Field Theory - Lattice 2012\\
June 24-29, 2012\\
Cairns, Australia}

\newcommand{\dslash}{\partial \hspace{-6.25pt}\slash}

\newcommand{\be}{\begin{equation}}
\newcommand{\ee}{\end{equation}}

\newcommand{\beann}{\begin{eqnarray*}} 
\newcommand{\eeann}{\end{eqnarray*}}

\begin{document}

\section{Motivation and overview}
Despite the fact that the latest results from the Large Hadron
Collider make it more and more unlikely that supersymmetry, at least
in its variety as a minimal extension to the Standard Model, is
accommodated in nature, supersymmetric quantum field theories remain
to be interesting in their own right. In particular, spontaneous
supersymmetry breaking and the corresponding phase transition is an
interesting non-perturbative phenonemon which often evades a
quantitative description even in simple models such as the ${\cal
  N}=1$ Wess-Zumino model in two dimension on which we focus in these
proceedings.  Often, a specific model may or may not undergo a
supersymmetry breaking phase transition and it is usually not clear
how such a transition is realised in detail. While the lattice
regularisation provides a convenient setup to perform detailed
non-perturbative numerical investigations, for systems which exhibit
spontaneous supersymmetry breaking straightforward Monte Carlo
simulations are not possible due to a fermion sign problem related to
the vanishing of the Witten index \cite{Baumgartner:2011cm}. However,
it has been shown that the sign problem can be circumvented by using
the fermion loop formulation
\cite{Baumgartner:2011cm,Baumgartner:2011jw,Baumgartner:2012np} and simulating the system
with the open fermion string algorithm \cite{Wenger:2008tq,Wenger:2009mi}.

In these proceedings, we present a quantitative non-perturbative
investigation of the $2d$ ${\cal N}=1$ Wess-Zumino model as
follows. First we give a brief definition of the model and then
discuss its formulation in terms of fermion loops. After reviewing its
vacuum structure and the symmetry breaking pattern we go on to
describe quantitatively its mass spectrum in the supersymmetric and
the supersymmetry broken phase as well as across the phase transition.

\section{The  ${\cal  N}=1$ Wess-Zumino model on the lattice}
The ${\cal N}=1$ Wess-Zumino model in two dimensions
\cite{Ferrara:1975nf} is one of the simplest models which may exhibit
spontaneous supersymmetry breaking.  Its degrees of freedom consist of
one real Majorana fermion field $\psi$ and one real bosonic field
$\phi$, while its dynamics is described by the Lagrangian density
\be
\label{eq:WZLagrangian}
{\cal L} = \frac{1}{2} \left(\partial_\mu \phi \right)^2
+ \frac{1}{2} P'(\phi)^2 
+ \frac{1}{2}\overline{\psi} \left( \dslash + P''(\phi) \right) \psi
\, .
\ee
Here, $P(\phi)$ denotes a generic superpotential, and $P', P''$ its
first and second derivative with respect to $\phi$. In the following
we will concentrate on the specific form  
\be
\label{eq:cubic superpotential}
P(\phi) = \frac{m^2}{4 g} \, \phi + \frac{1}{3} g \phi^3
\ee
which leads to a vanishing Witten index $W=0$ and hence allows for
spontaneous supersymmetry breaking \cite{Witten:1982df}.  The
corresponding action enjoys the following two symmetries. First, there
is a single supersymmetry given by the transformations
\be
\delta \phi = \overline{\epsilon} \psi \, , \quad 
\delta \psi = (\dslash \phi - P') \epsilon \, , \quad
\delta \overline{\psi} = 0 \, , 
\ee
and secondly, there is a discrete  $\mathbb{Z}(2)$ chiral symmetry given
by
\be
\phi \rightarrow  -\phi \, , \quad
\psi \rightarrow \gamma_5 \psi  \, , \quad 
\overline{\psi} \rightarrow  - \overline{\psi} \gamma_5 \, ,
\ee
where $\gamma_5 \equiv \sigma_3$ can be chosen to be the third Pauli
matrix.  The fact that the Witten index is zero for the chosen
superpotential can be derived from the transformation properties of
the Pfaffian of the Dirac operator under the $\mathbb{Z}(2)$ symmetry
$\phi \rightarrow -\phi$ \cite{Baumgartner:2012np}.

Let us now move on to describe the regularisation of the model on the
lattice.  For the fermionic fields we use the Wilson lattice
discretisation yielding the fermion Lagrangian density
\[
 {\cal L} = \frac{1}{2} \xi^T {\cal C} (\gamma_\mu \tilde \partial_\mu
 - \frac{1}{2} \partial^* \partial + P''(\phi)) \xi \, ,
\]
where $\xi$ is a real, 2-component Grassmann field, ${\cal C} = -{\cal
  C}^T$ is the charge conjugation matrix and $\partial^*, \partial$
are the backward and forward lattice derivatives, respectively. In
order to guarantee the full supersymmetry in the continuum limit, one
needs to introduce the same derivative, in particular the Wilson term,
also for the bosonic fields \cite{Golterman:1988ta}. As a consequence,
in addition to the supersymmetry also the $\mathbb{Z}(2)$ chiral
symmetry is broken by the lattice regularisation both in the bosonic
and the fermionic sector.

Nevertheless we can now use the exact reformulation of the fermionic
degrees of freedom in term of closed fermion loops
(cf.~\cite{Baumgartner:2011cm} for further details). Together with the
fermion string algorithm \cite{Wenger:2008tq,Wenger:2009mi} this
allows simulations with unspecified fermionic boundary conditions
which do not suffer from the fermion sign problem
\cite{Baumgartner:2012np} and for which critical slowing down is
essentially absent even in the presence of a massless fermionic mode
such as the Goldstino.

\section{Supersymmetry breaking pattern}
It is useful to briefly review the (super-)symmetry breaking pattern. The
potential for the bosonic field is a standard $\phi^2$-theory which
may trigger a $\mathbb{Z}(2)$ symmetry breaking phase transition. In
particular, for large $m/g$ one expects that the $\mathbb{Z}(2)$
symmetry is broken. In that case, the vacuum expectation value of the
boson field $\langle \overline \phi\rangle = \pm m/2g$ is expected to
select a definite ground state for the system, either bosonic or
fermionic. On the other hand, for small $m/g$ one expects the
$\mathbb{Z}(2)$ symmetry to be restored with $\langle \overline \phi
\rangle = 0$ in which case no unique ground state is selected and
hence supersymmetry is broken. In fact, the associated tunneling
between the two allowed bosonic and fermionic vacua corresponds to the
infamous massless Goldstino mode.

In \cite{Baumgartner:2012np} it was indeed demonstrated, using the
Witten index
\[
W \equiv Z_\text{pp} = Z_{{\cal L}_{00}} - Z_{{\cal L}_{10}} - Z_{{\cal L}_{01}} -
Z_{{\cal L}_{11}} \, ,
\]
as an order parameter, that a supersymmetry breaking phase transition
occurs for specific couplings $\hat g/\hat m$ depending on the lattice
spacing set by $a g$. Here, $Z_\text{pp}$ denotes the partition
function with periodic boundary conditions in both directions while
$Z_{{\cal L}_{ij}}$ denote partition functions with fixed topological
boundary conditions \cite{Baumgartner:2011jw}.
The expected symmetry breaking pattern and the corresponding vacuum
structure follow exactly the expectations described above. In
particular, for large $m/g$ one is in a $\mathbb{Z}(2)$ broken phase
where supersymmetry is unbroken, while for small $m/g$ the
$\mathbb{Z}(2)$ symmetry is restored and the supersymmetry is
broken. Note that this situation only holds in the infinite volume
limit: at any finite volume the $\mathbb{Z}(2)$ symmetry is always
restored (and hence the supersymmetry broken) by soliton solutions
which mediate transitions between boson field configurations with
$\langle \overline \phi\rangle = \pm m/2g$ \cite{Catterall:2003ae}. We
have now further confirmed this scenario using the Ward identity
$\langle P' \rangle$.

\begin{figure}[t]
\includegraphics[width=0.5\textwidth]{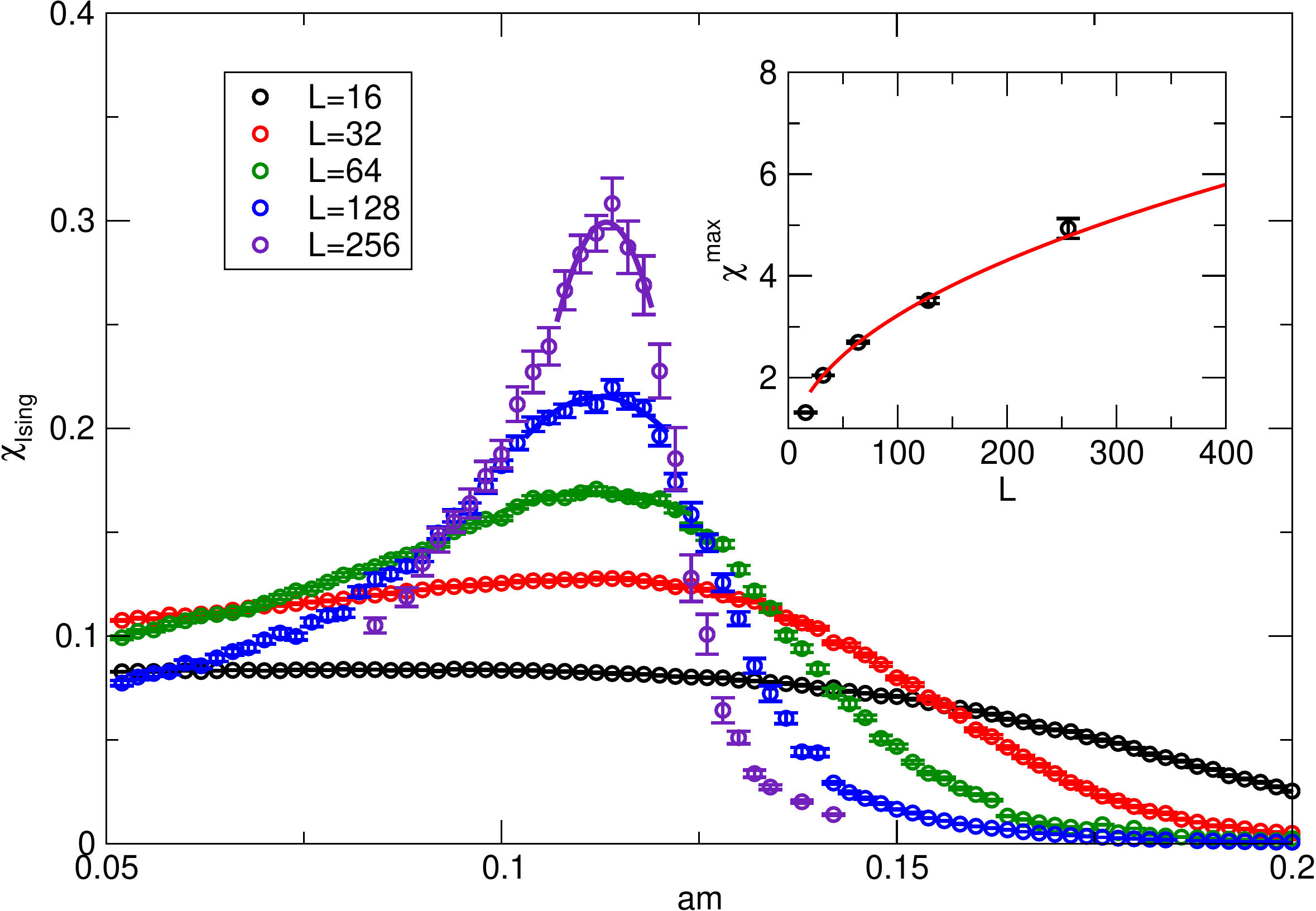} 
\includegraphics[width=0.49\textwidth]{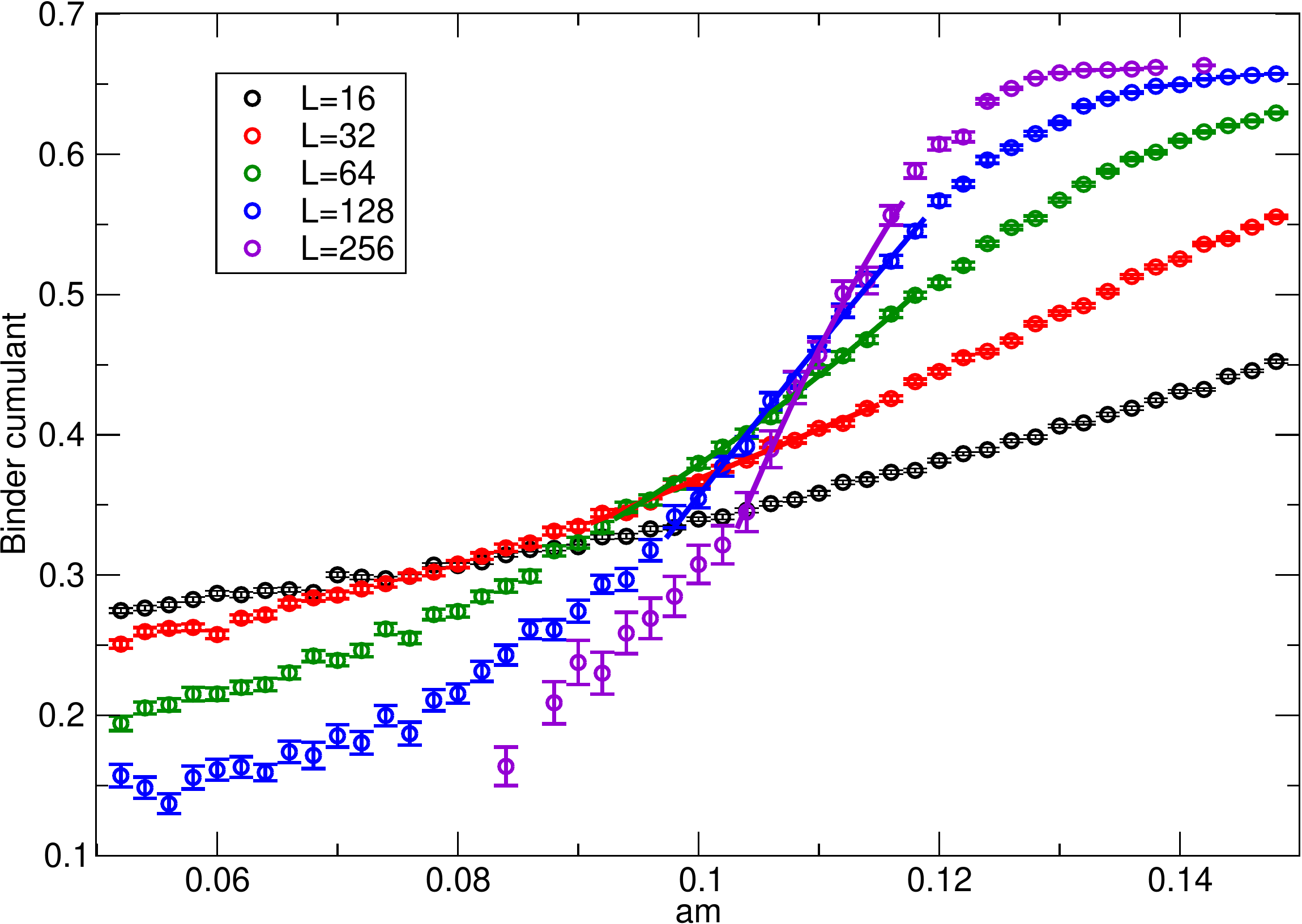} 
\caption{Susceptibility of the volume averaged Ising projected boson
  field (left plot) and the Binder cumulant of the boson field (right
  plot) for several volumes at $a g =
  0.0625$. 
\vspace{-0.2cm} 
}
\label{fig:ising_susceptibility}
\end{figure}

\subsection{$\mathbb{Z}(2)$ phase transition}
In order to further quantify the phase transition we investigated in
detail several order parameters sensitive to the $\mathbb{Z}(2)$ phase
transition. It should be noted that there exists no real order
parameter for the $\mathbb{Z}(2)$ transition, since the Wilson lattice
discretisation breaks not just the supersymmetry, but also the
$\mathbb{Z}(2)$ chiral symmetry in the bosonic sector. However, it
turns out that at the lattice spacings $ag \le 0.25$ which we
simulated, the system behaves sufficiently continuum-like, so that
accurate determinations of the phase transition are possible without
problems. This is exemplified in figure
\ref{fig:ising_susceptibility}.  In the left plot we show the
susceptibility $\chi_\text{Ising}$ of the volume averaged Ising
projected boson field $\overline{\phi}_\text{Ising} = 1/V \sum_x
\text{sign}[\phi_x]$. The susceptibility shows a nice finite volume
scaling and the scaling of the susceptibility peak indicates a second
order phase transition, presumably in the universality class of the
$2d$ Ising model. The right plot of figure
\ref{fig:ising_susceptibility} shows the Binder cumulant of the boson
field for various volumes, all at fixed lattice spacing $ag =
0.0625$. From the position of the susceptibility peak and the crossing
of the Binder cumulant one can infer the critical bare mass $am_c$ at
which the phase transition occurs.

\begin{figure}[t]
\includegraphics[width=0.5\textwidth]{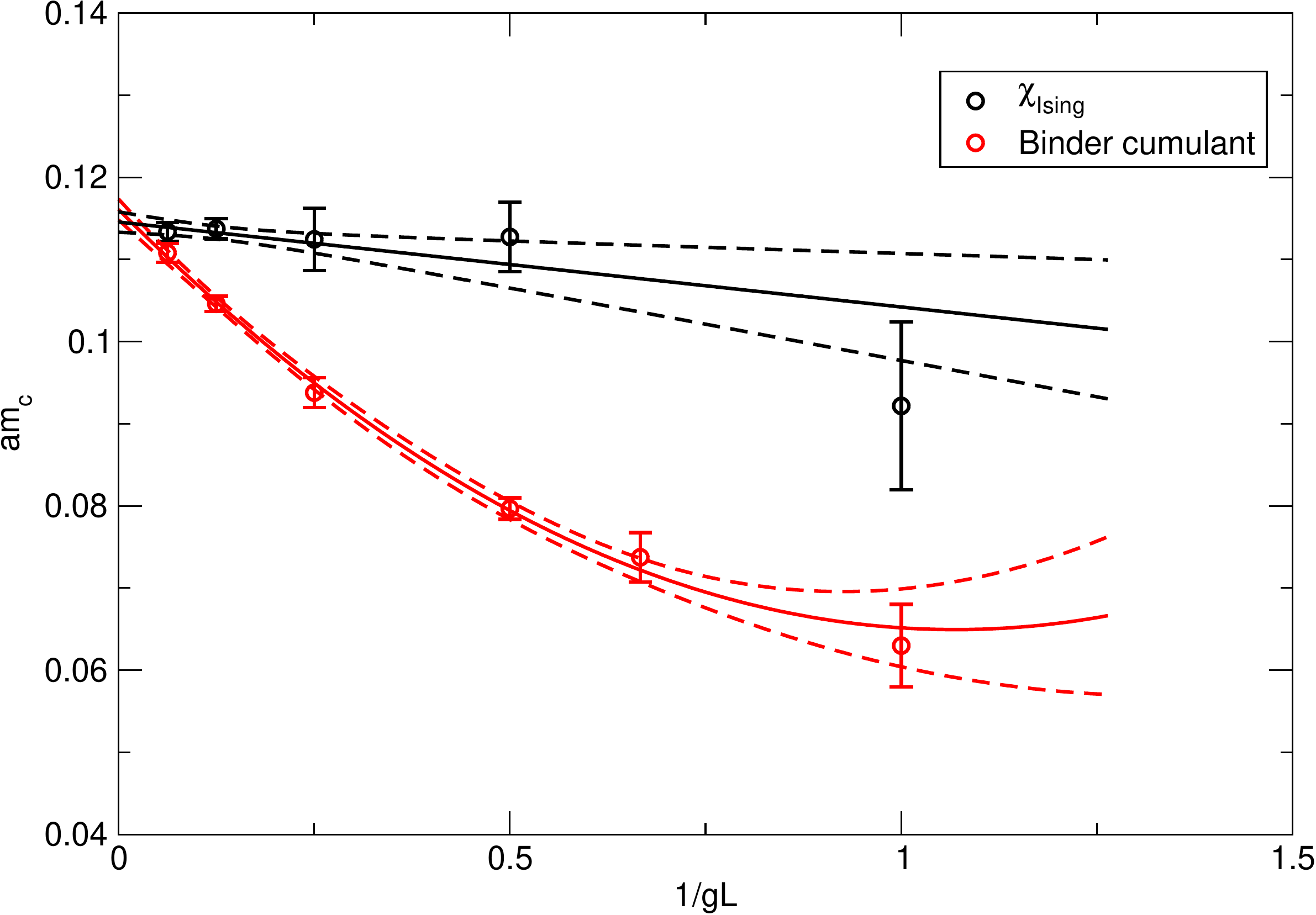} 
\includegraphics[width=0.5\textwidth]{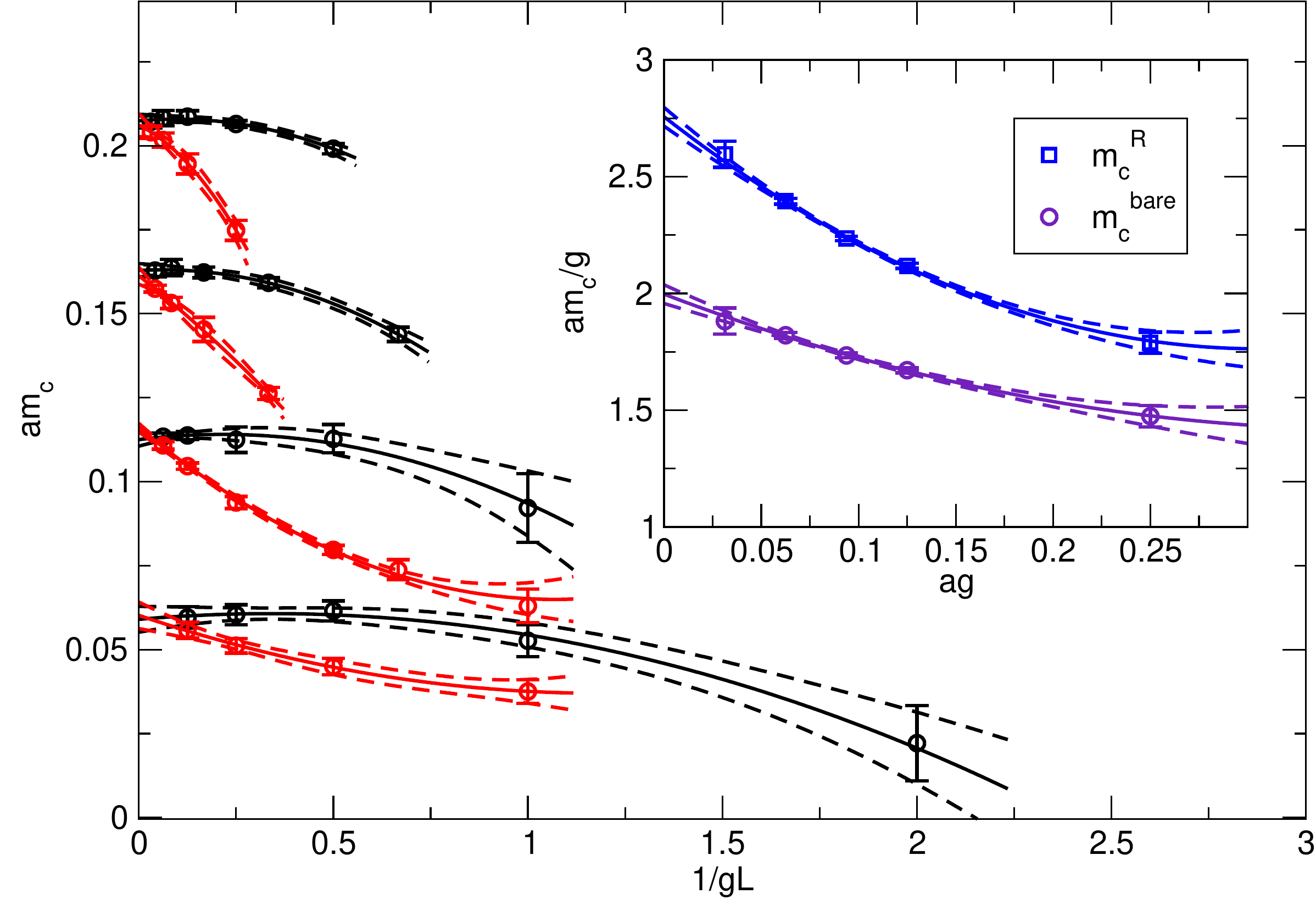} 
\caption{Thermodynamic limit of the critical mass $am_c$ from the
  Binder cumulant and the peak of the susceptibility at $ag=0.0625$
  (left plot) and for a range of other couplings
  $ag=0.25-0.03125$ (right plot). The inset shows the continuum limit of the bare
  and the renormalised critical coupling $f_c=g/m_c^\text{bare,R}$.
\vspace{-0.4cm} 
}
\label{fig:thermodynamicLimit}
\end{figure}

In general, different order parameters consistently indicate a phase
transition only in the thermodynamic limit when the finite volume
pseudo-phase transition becomes a true one. In the left plot of figure
\ref{fig:thermodynamicLimit} we show the critical bare mass $am_c$ as
a function of the inverse volume expressed in units of $g$, as
obtained from the two (pseudo-)order parameters discussed above. We
find that the determination from the Binder cumulant shows rather
large finite size effects, in contrast to the one from the
susceptibility. However, in the thermodynamic limit they both agree
and this is sustained for all lattice spacings (right plot). The inset
finally shows the continuum extrapolation of the critical coupling
$f_c= g/m_c$ using the bare mass $am_c$ and the one renormalised using
1-loop continuum perturbation theory, $am_c^\text{R}$. The
renormalised critical coupling in the continuum can now be compared to
the one obtained in \cite{Wozar:2011gu} using a different
discretisation and algorithm.

\section{Mass spectrum}
We determine the mass spectrum from the temporal behaviour of
correlators projected to zero spatial momentum, $C(t) \sim \langle
{\cal O}(0) {\cal O}^{(T)}(t)\rangle$. For the boson masses we use the
$\mathbb{Z}(2)$-odd and -even operators ${\cal O} = \phi$ and
$\phi^2$, respectively, while for the fermion masses we use ${\cal O}
= \xi$ and $\xi \phi$. We note that in the
supersymmetric/$\mathbb{Z}(2)$-broken phase the vacuum can not
distinguish between the even and odd states and hence we extract the
same mass using the two operators, while in the supersymmetry (SUSY)
broken/$\mathbb{Z}(2)$ restored phase, the vacuum respects the
$\mathbb{Z}(2)$ symmetry and distinguishes between states with
different $\mathbb{Z}(2)$ quantum numbers. Furthermore, in the SUSY
broken phase we can measure excitations both in the bosonic vacuum,
i.e.~in $Z_{{\cal L}_{00}}$, and in the fermionic one, i.e.~$Z_{{\cal
    L}_{10}} + Z_{{\cal L}_{01}} + Z_{{\cal L}_{11}}$. We emphasise
that simulations in the SUSY broken phase are only feasible due to the
fact that the fermion loop algorithm essentially eliminates critical
slowing down \cite{Wenger:2008tq,Wenger:2009mi}, despite the emergence of the
(would-be) Goldstino.

\begin{figure}[t]
\includegraphics[width=0.5\textwidth]{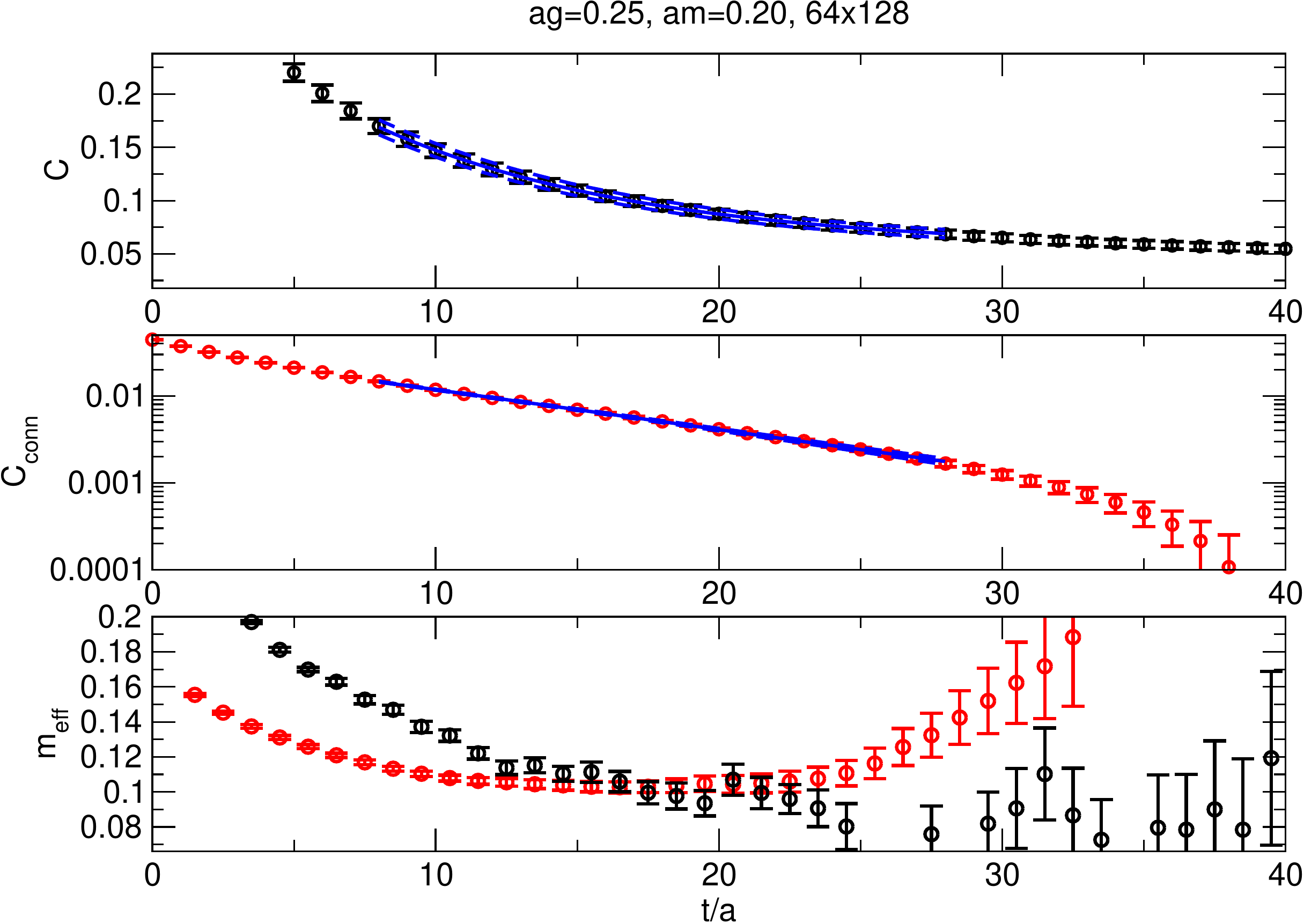} 
\includegraphics[width=0.5\textwidth]{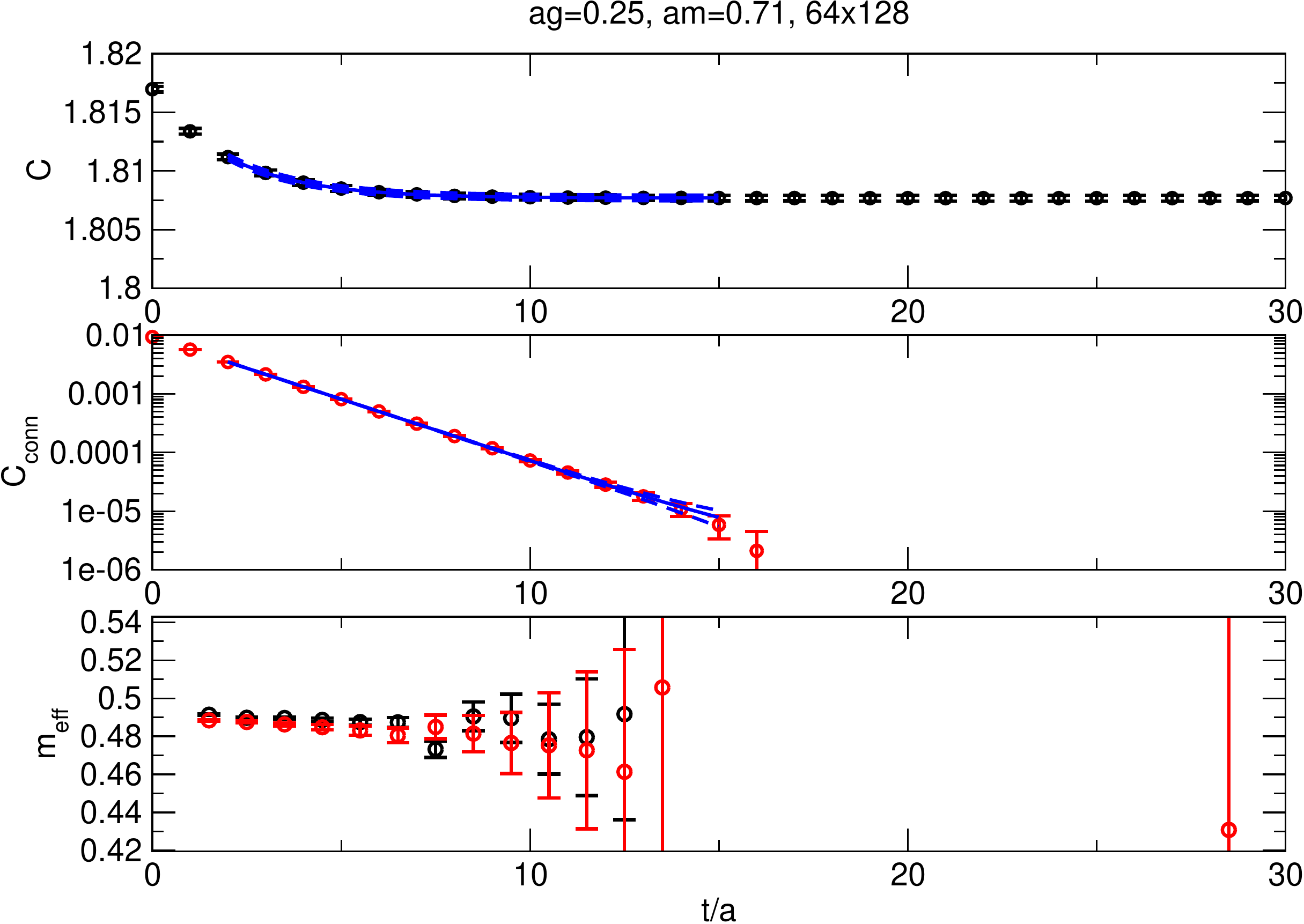} 
\caption{Boson mass extraction in the SUSY
  broken/$\mathbb{Z}(2)$-symmetric (left plot) and in the supersymmetric/$\mathbb{Z}(2)$ broken phase (right plot).
\vspace{-0.2cm} 
}
\label{fig:mass_extraction_boson}
\end{figure}

In figure \ref{fig:mass_extraction_boson} we show examples of boson
mass extractions in the SUSY broken/$\mathbb{Z}(2)$-symmetric 
(left plot) and in the supersymmetric/$\mathbb{Z}(2)$ broken phase
(right plot), both in the bosonic vacuum. The top panel shows the full
correlator, the middle one the connected part and the lowest one
 the corresponding effective masses. In the SUSY broken phase we can
 fit double exponentials (plus a small shift due to the residual
 $\mathbb{Z}(2)$ breaking), while in the $\mathbb{Z}(2)$ broken phase
 only one exponential can be fitted, since the signal is quickly
 dominated by the fluctuations stemming from the large disconnected contribution.

\begin{figure}[b]
\includegraphics[width=0.5\textwidth]{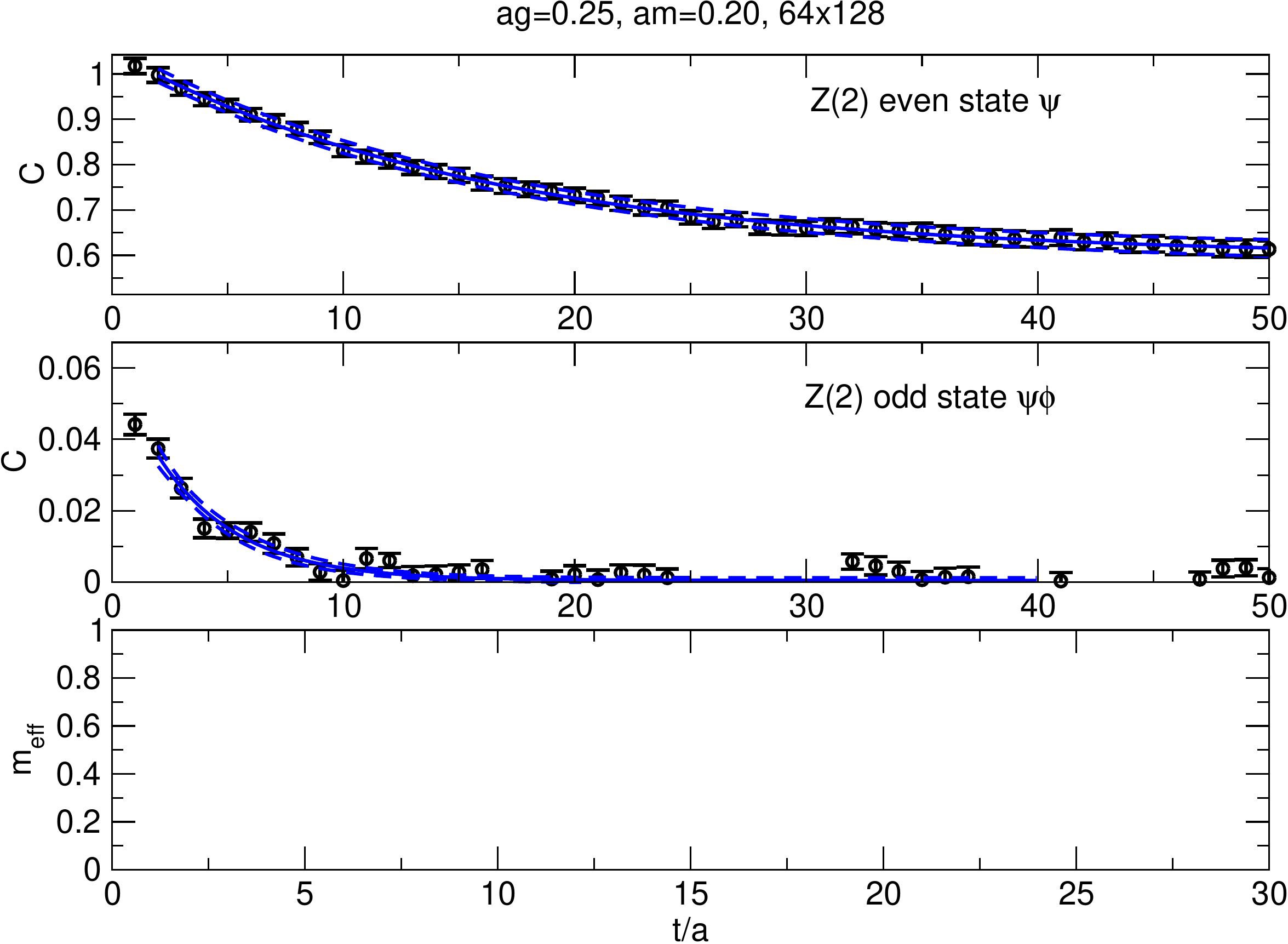} 
\includegraphics[width=0.5\textwidth]{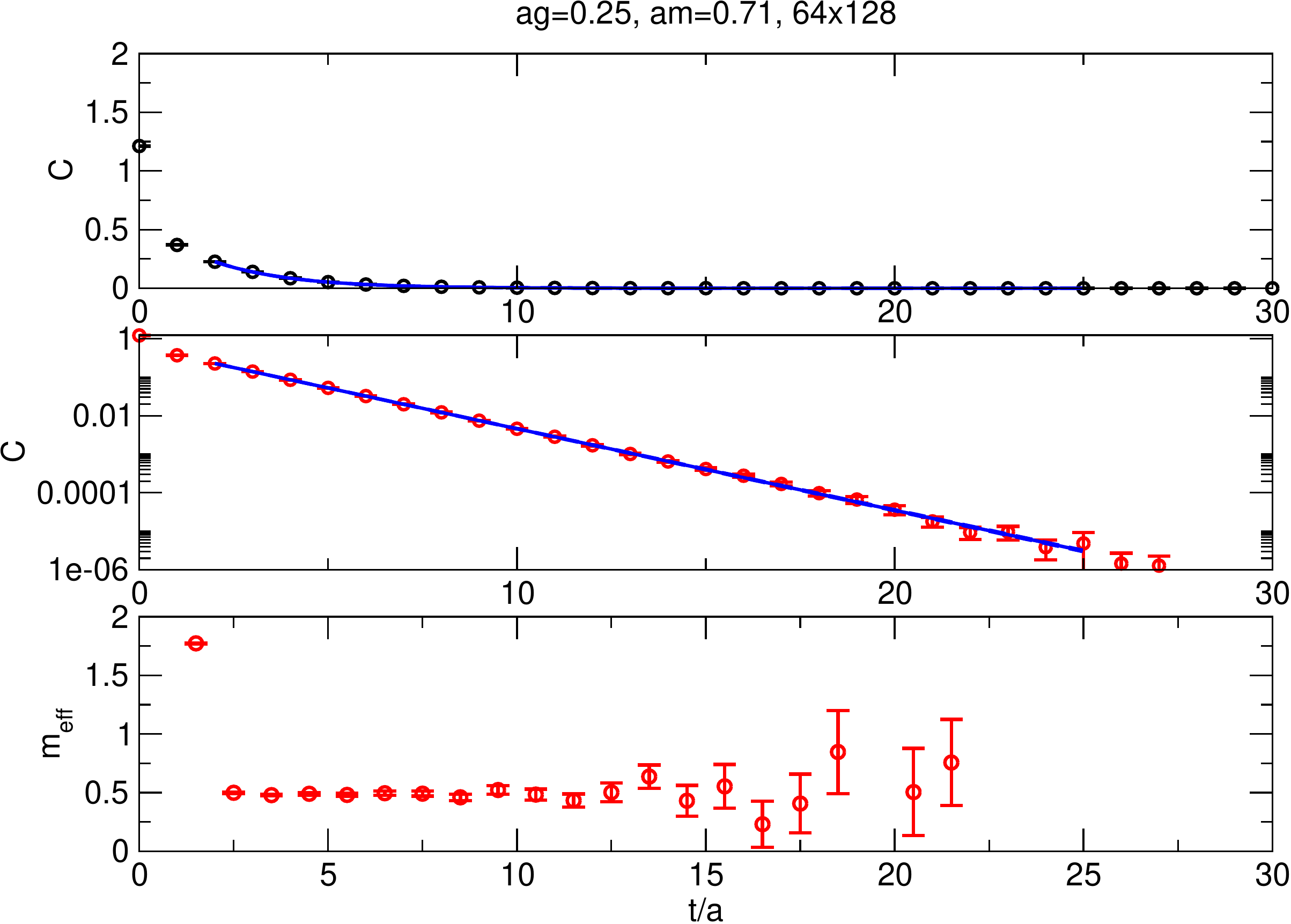} 
\caption{Fermion mass extraction in the SUSY
  broken/$\mathbb{Z}(2)$-symmetric (left plot) and in the supersymmetric/$\mathbb{Z}(2)$ broken phase (right plot).
\vspace{-0.2cm} 
}
\label{fig:mass_extraction_fermion}
\end{figure}

In figure \ref{fig:mass_extraction_fermion} we show examples of
fermion mass extractions in both phases. In the left plot (SUSY broken
phase) the top panel shows the correlator of the $\mathbb{Z}(2)$-even
state which can be well fitted with a double exponential with the
lowest mass corresponding to the Goldstino mass. The middle panel
shows the $\mathbb{Z}(2)$-odd state fitted with a single
exponential. The right plot shows the fermion correlator in the
supersymmetric phase (top panel), on a log scale (middle panel) and
the corresponding effective masses (bottom panel). It is remarkable
that the signal of the fermion correlator can be followed over more
than six orders of magnitude. Of course this just reflects the
efficiency of the employed fermion loop algorithm
\cite{Wenger:2008tq}.

\begin{figure}[t]
\includegraphics[width=0.5\textwidth]{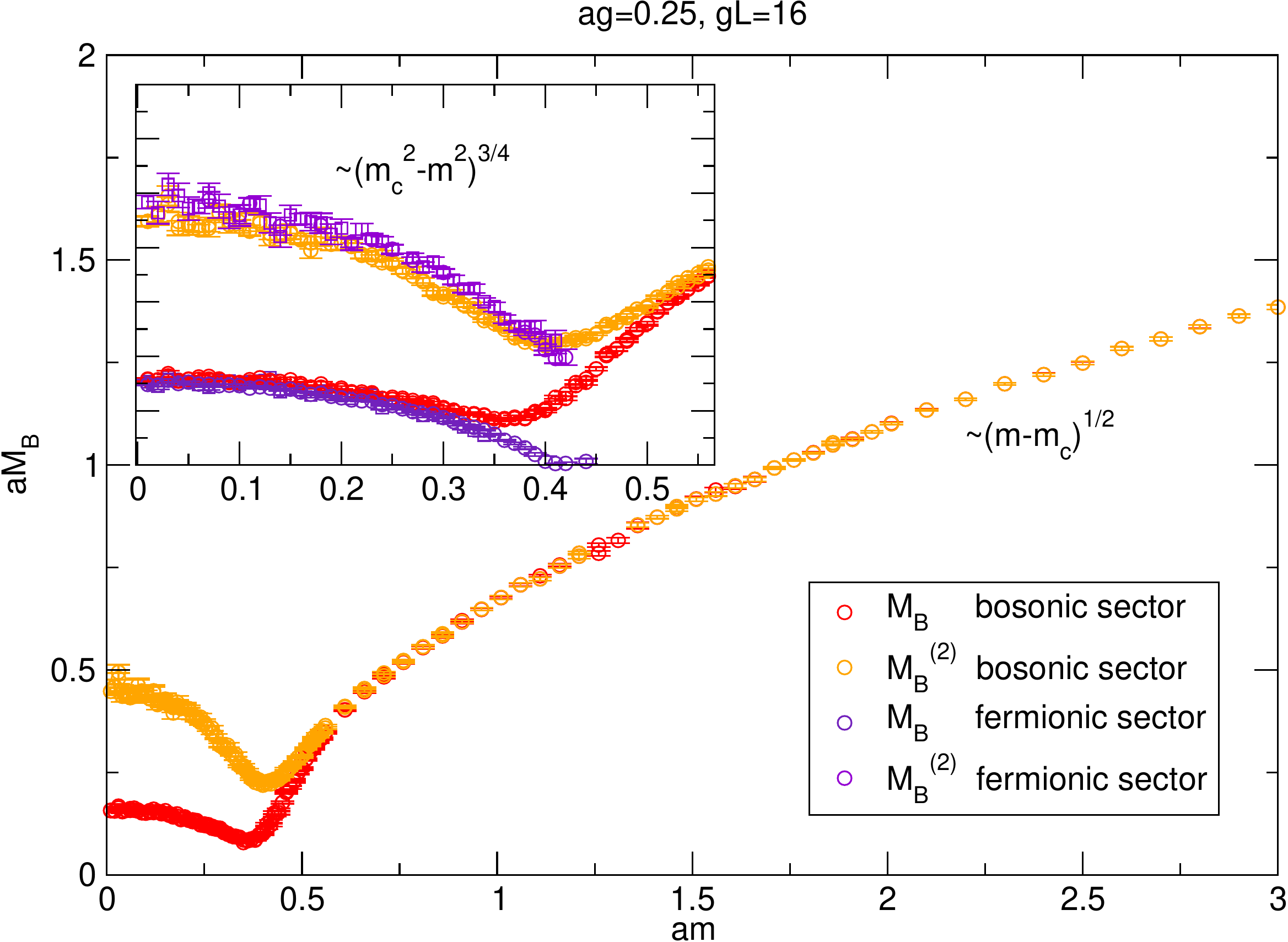} 
\includegraphics[width=0.5\textwidth]{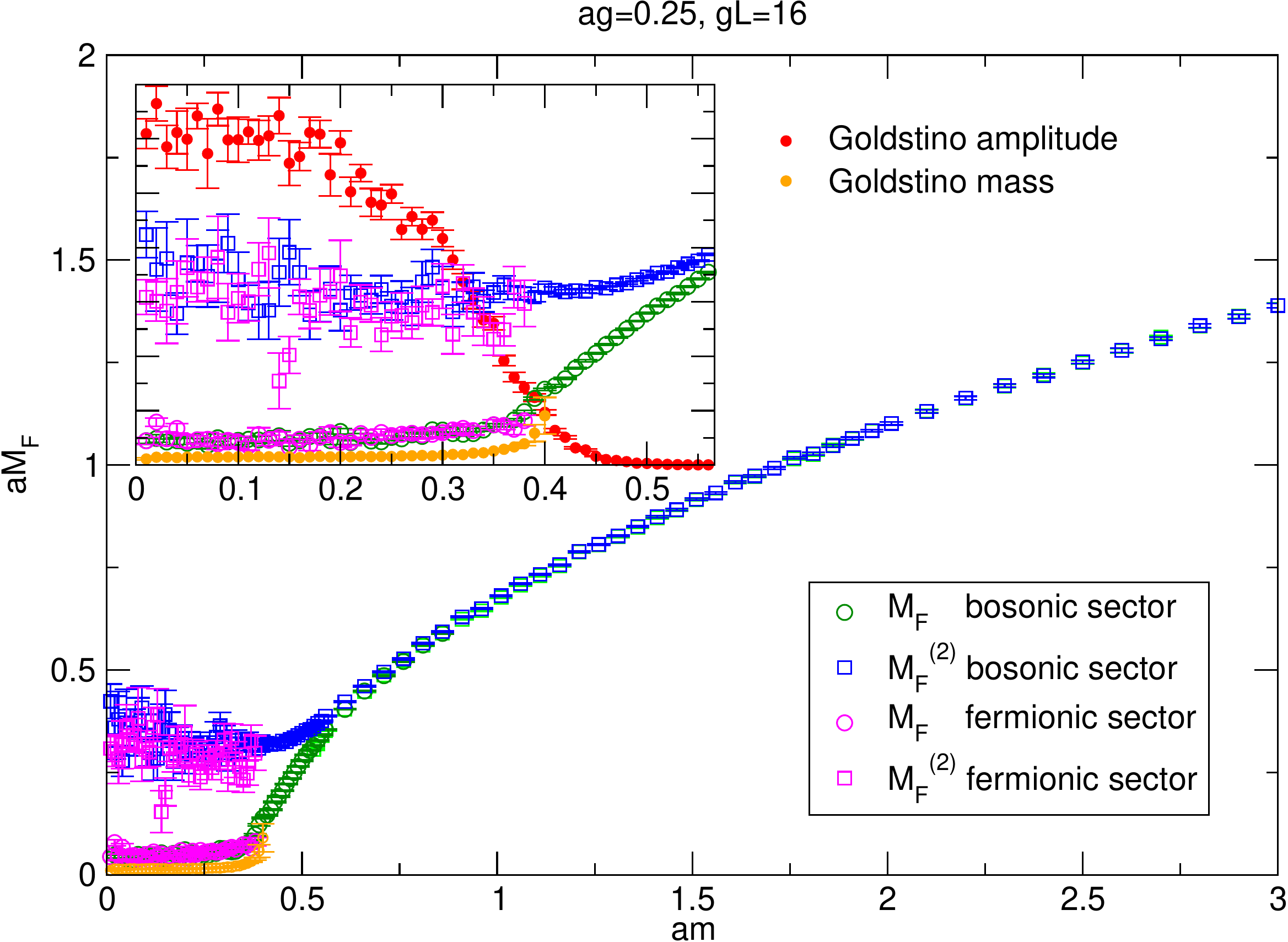} 
\caption{Mass spectrum for bosonic (left plot) and
  fermionic excitations (right plot). The superscript ${}^{(2)}$
  denotes the excited state.
\vspace{-0.2cm} 
}
\label{fig:mass_spectrum}
\end{figure}

Finally, in figure \ref{fig:mass_spectrum} we show the full boson and
fermion mass spectrum in the left and right plot, respectively, as a
function of the bare mass $am$ across the supersymmetry breaking phase
transition occurring at around $am_c \sim 0.042$. We see how the mass
spectrum in the SUSY broken/$\mathbb{Z}(2)$-symmetric phase fans out
into the $\mathbb{Z}(2)$-even and -odd states, with bosonic and
fermionic masses non-degenerate, while in the
supersymmetric/$\mathbb{Z}(2)$-broken phase the states collapse onto a
degenerate mass, in addition to the boson and fermion masses being
equal.  In the SUSY broken phase we can crosscheck the mass
determination in the bosonic sector with the one in the fermionic
sector and we find very convincing consistency. This agreement in the
SUSY broken phase and the degeneracy of the boson and fermion masses
in the supersymmetric phase is rather surprising, given the fact that
the simulations are at finite and rather coarse lattice spacing
$ag=0.25$. Moreover, it should be kept in mind, that in the SUSY
broken phase it is rather difficult to keep the systematic effects
from mixing with higher excited states under control.

A first preliminary investigation of the effects of the finite volume
on the spectrum reveals that they are essentially negligible for the
volume $L/a =64$ that we are using here. This is not quite the case
for the boson mass spectrum in the  SUSY broken phase. In fact, the
investigation in \cite{Synatschke:2009nm} suggests a distinct finite
volume scaling of the boson masses with the lowest boson mass
vanishing towards the thermodynamic limit.

An interesting feature of the spectrum of a theory with spontaneously
broken supersymmetry is of course the occurrence of the massless
Goldstino. Since in our regularisation the supersymmetry is broken
explicitely at any finite lattice spacing, the Goldstino is only
approximately massless as can be seen in figure
\ref{fig:mass_spectrum}. To corroborate the identification of this low
mass state as the Goldstino, we plot in the inset of the right plot
also the contribution (amplitude) of that state to the full fermion
correlator. It turns out that the amplitude decreases as we increase
the bare mass and vanishes at the transition to the supersymmetric
phase, i.e.~the Goldstino decouples from the system at the
supersymmetry restoring phase transition.

\bibliographystyle{JHEP}
\bibliography{ssb2dWZm}

\end{document}